\documentclass[preprint2]{aastex}

\shorttitle{BINARY FREQUENCY OF $r$-PROCESS-ELEMENT ENHANCED METAL-POOR STARS}
\shortauthors{Hansen et al.}

\begin{document}

\title{THE BINARY FREQUENCY OF $r$-PROCESS-ELEMENT ENHANCED METAL-POOR STARS AND
ITS IMPLICATIONS: CHEMICAL TAGGING IN THE\\ PRIMITIVE HALO OF THE MILKY WAY}

\author{Terese Hansen, Johannes Andersen\altaffilmark{1}, Birgitta Nordstr\"om,
Lars A. Buchhave\altaffilmark{2}} \affil{The Niels Bohr Institute,
University of Copenhagen, Juliane Maries Vej 30, DK-2100 Copenhagen,
Denmark} \altaffiltext{1}{Also at Nordic Optical Telescope Scentific
Association, La Palma, Spain} \altaffiltext{2}{Also at the Centre
for Star and Planet Formation, Natural History Museum of Denmark,
University of Copenhagen, DK-1350 Copenhagen, Denmark}
\email{terese@fys.ku.dk, ja@astro.ku.dk, birgitta@astro.ku.dk,
buchhave@astro.ku.dk} \and
\author{Timothy C. Beers}
\affil{Department of Physics \& Astronomy and JINA: Joint Institute for Nuclear Astrophysics,\\
Michigan State University, East Lansing, MI 48824, USA}
\email{beers@pa.msu.edu}

\begin{abstract}

A few rare halo giants in the range [Fe/H] $\simeq -2.9\pm0.3$ exhibit {\it
r}-process element abundances that vary as a group by factors up to
[$r$/Fe] $\sim80$, relative to those of the iron peak and below. Yet, the
astrophysical production site of these {\it r}-process elements remains
unclear. We report initial results from four years of monitoring the radial
velocities of 17 {\it r}-process-enhanced metal-poor giants to detect and
characterise binaries in this sample. We find three (possibly four)
spectroscopic binaries with orbital periods and eccentricities that are
indistinguishable from those of Population I binaries with giant primaries,
and which exhibit no signs that the secondary components have passed through the
AGB stage of evolution or exploded as supernovae. The other 14 stars in our
sample appear to be single -- including the prototypical
$r$-process-element enhanced star CS~22892-052, which is
also enhanced in carbon, but not in {\it s}-process elements.\\ We conclude
that the {\it r}-process (and potentially carbon) enhancement of these
stars was not a local event due to mass transfer or winds from a binary
companion, but was imprinted on the natal molecular clouds of these
(single and binary) stars by an external source. These stars are thus
spectacular chemical tracers of the inhomogeneous nature of the early
Galactic halo system.

\end{abstract}

\keywords{Galaxy: halo --- Stars: Population II  --- binaries: spectroscopic --- techniques: radial velocities}


\section{INTRODUCTION}

The chemical composition of very metal-poor (VMP; [Fe/H] $< -2.0$) and
extremely metal-poor (EMP; [Fe/H] $< -3.0$) stars provides a fossil record
of the star formation and nucleosynthesis history of the early Galaxy.
Carbon (and often N and O) enhancement appears to be common for stars with
[Fe/H] $\lesssim -2.5$ \citep{BeCh05,Carol11}, but the overall abundance
ratios of elements up to the iron peak are well established by [Fe/H]
$\simeq -2.5$, with very small scatter \citep{LP5,Arn05}. The full range of
{\it r}-process elements was also in place at this stage in the chemical
evolution of the Galaxy \citep[see the review by][]{sneden08}. Other
neutron-capture processes, such as the {\it s-}process, began to contribute
significant amounts of heavy elements from [Fe/H] $\lesssim -2.5$
\citep[e.g, ][]{sim04}\footnote{Note that \citet{roederer10} have argued
that the {\it majority} of $s-$process element enhancement may have been
delayed until [Fe/H] $\sim -1.4$. See also the discussion in
\citet{Bisterzo11}.}. Hence, the processes by which the chemical elements
were produced and recycled into the Galactic halo system are, at least in a
broad-brush sense, well understood.

However, in a small fraction of stars in the abundance range $-3.2 <$
[Fe/H] $< -2.6$, the uniform abundance pattern of the {\it r}-process
elements as a group is enhanced by factors up to $\sim80$ relative to that
of the iron-peak and lighter elements. Spectroscopic analyses of such stars
with 8m-class telescopes have provided precise and detailed abundances of
many {\it r}-process elements as a key to understanding their origin
\citep{sneden08, cowan11}. Yet, the likely production site(s) of the {\it
r}-process elements, as well as the mechanisms by which their abundances
relative to the ``standard'' halo composition could vary so strongly from
star to star in the early Galaxy, remain essentially unknown.

Explanations of this diversity fall into two main classes: Inhomogeneous
enrichment and incomplete mixing of the ISM by the first generation(s) of
stars, or later, local pollution of neutron-capture elements by a binary
companion of the presently observed star \citep{QW01}. In the first case,
the very uniform abundance pattern of the $\alpha$- and iron-peak elements
is difficult to reconcile with the predictions of models of stochastic star
formation and enrichment in the early Galaxy \citep{Arn05}. This conflict
would be resolved in the second case, but unlike the {\it
s}-process-element enhanced Ba and CH giants, which are known to {\it all} be
long-period binaries \citep{McCW90, Joris98}, this conjecture is so far
without observational foundation.

Here we report the first results from four years of precise radial-velocity
monitoring, performed in order to assess the binary frequency of a sample
of 17 {\it r}-process-element enhanced VMP and EMP giants. Our results
provide strong new constraints on the nature of the {\it r}-process
production site(s) and on the use of these stars as tracers of the star
formation and/or merger history of the early Galaxy.

\section{Sample Definition and Observations}
\label{sect_obs}

Our program stars were drawn from the HERES search for {\it
r}-process-element enhanced stars \citep{Chri04,Barklem05}, supplemented by
earlier and later discoveries as summarised by \citet{Hayek09}. Only stars
north of declination $\sim -25\degr$ and brighter than $V \sim 16.0$ are
accessible for study with the Nordic Optical Telescope (NOT) on La Palma,
resulting in a total sample of 17 stars. Eight of these are in the r-I
class ($+0.3 <$ [Eu/Fe] $<+1.0$), as defined by \citet{BeCh05}, and 9 are
in the r-II class ([Eu/Fe] $>+1.0$). Table \ref{tbl-1} lists the program
stars, in order of increasing [$r$/Fe] ratio, in order to highlight the
{\it continuum} of {\it r}-process enhancements that exists; the division
into r-I and r-II classes appears to be merely one of
convenience\footnote{Note that the r-II and r-I stars do occupy different,
but overlapping, regions of metallicity space. The r-II stars are found
{\it only} at very low metallicity, while the r-I stars extend over the
range $-3.0 \leq$ [Fe/H] $\leq -0.5$; see the bottom panel of Fig. 3 in
\citet{aoki10}.}. Stars with measured U abundances, and those found here to
be spectroscopic binaries, are indicated in the Table.

As the chemical compositions of the targets are already known -- some in
great detail -- our observations were designed to just yield precise radial
velocities as efficiently as possible, over a time span of several years.
To this end, we obtained high-resolution spectra ($R \sim 45,000$) with a
S/N ratio $\sim 10$, using the bench-mounted, fibre-fed \'{e}chelle
spectrograph FIES at the NOT \citep{ADJA10}, which is installed in a
separate, temperature-controlled underground enclosure. The useful
wavelength range covered by these spectra is $4000-7000$ \AA.

Our goal was to reach an accuracy of $100-200$ m s$^{-1}$ per
observation, except perhaps for the faintest program stars. The
radial-velocity zero point was checked with standard stars on every
night of observation, and found to be reproducible to better than 45
m s$^{-1}$ per observation over a five-year period. Thus, the
accuracy of the radial velocities of the program stars is not
limited by the instrument.

Our initial assumption, by analogy with the Ba and CH giants
\citep{Joris98}, was that any spectroscopic binaries in the sample
would likely have orbits of long period, low eccentricity, and small
amplitude -- i.e., small, slow velocity variations. Thus, our
strategy was to observe these stars at roughly monthly intervals,
weather permitting, and then adapt the frequency of the observations
to follow any objects with radial-velocity variations detected in
the initial data. Observations began in April 2007, and were
continued on 51 nights through September 2011, for a total of $\sim$
234 spectra, an average of 14 per star. Faint stars at far southern
declinations require ideal conditions, and were observed less
frequently than average.

\begin{deluxetable}{lrrrrrrrl}
\tabletypesize{\scriptsize}
\tablecaption{Stars Monitored for Radial Velocity Variation \label{tbl-1}}
\tablewidth{0pt}
\tablehead{
\colhead{Star} & \colhead{RA (J2000)} & \colhead{Dec (J2000)} &
\colhead{$V$} & \colhead{$B-V$} & \colhead{[Fe/H]} & \colhead{[r/Fe]}
& \colhead{Nobs} & \colhead{Remarks}
}
\startdata
HE 0524-2055 & 05:27:04 & $-$20:52:42 & 14.01 & 0.87  & $-$2.58 & $+$0.49 &  8 & r-I  \\
HE 0442-1234 & 04:44:52 & $-$12:28:46 & 12.91 & 1.07  & $-$2.41 & $+$0.52 & 23 & r-I, SB \\
HE 1430+0053 & 14:33:17 & $+$00:40:49 & 13.69 & 0.58  & $-$3.03 & $+$0.72 & 19 & r-I  \\
CS 30315-029 & 23:34:27 & $-$26:42:19 & 13.66 & 0.91  & $-$3.33 & $+$0.72 &  9 & r-I  \\
HD 20        & 00:05:15 & $-$27:16:18 &  9.07 & 0.54  & $-$1.58 & $+$0.80 &  9 & r-I  \\
HD 221170    & 23:29:29 & $+$30:25:57 &  7.71 & 1.02  & $-$2.14 & $+$0.85 & 23 & r-I  \\
HE 1044-2509 & 10:47:16 & $-$25:25:17 & 14.34 & 0.66  & $-$2.89 & $+$0.94 & 13 & r-I, SB \\
HE 2244-1503 & 22:47:26 & $-$14:47:30 & 15.30 & 0.60  & $-$2.88 & $+$0.95 & 12 & r-I  \\
HE 2224+0143 & 22:27:23 & $+$01:58:33 & 13.68 & 0.71  & $-$2.58 & $+$1.05 & 18 & r-II \\
HE 1127-1143 & 11:29:51 & $-$12:00:13 & 15.88 & \dots & $-$2.73 & $+$1.08 & 12 & r-II \\
HE 0432-0923 & 04:34:26 & $-$09:16:50 & 15.16 & 0.73  & $-$3.19 & $+$1.25 & 14 & r-II \\
HE 1219-0312 & 12:21:34 & $-$03:28:40 & 15.94 & 0.64  & $-$2.81 & $+$1.41 &  6 & r-II \\
CS 22892-052 & 22:17:01 & $-$16:39:26 & 13.21 & 0.80  & $-$2.95 & $+$1.54 & 17 & r-II \\
CS 29497-004 & 00:28:07 & $-$26:03:03 & 14.03 & 0.70  & $-$2.81 & $+$1.62 &  8 & r-II \\
CS 31082-001 & 01:29:31 & $-$16:00:48 & 11.66 & 0.76  & $-$2.78 & $+$1.66 & 17 & r-II, U \\
HE 1523-0901 & 15:26:01 & $-$09:11:38 & 11.10 & 1.10  & $-$2.95 & $+$1.80 & 17 & r-II, U, SB \\
HE 1105+0027 & 11:07:49 & $+$00:11:38 & 15.64 & 0.39  & $-$2.42 & $+$1.81 &  9 & r-II  \\
\enddata
\tablecomments{Remarks: U indicates that uranium has been detected; SB indicates a confirmed spectroscopic binary.}
\end{deluxetable}

\section{Data Reduction and Analysis}\label{sect_data}

The entire set of reductions of the raw spectra (bias subtraction, division
by a flat-field exposure, cosmic ray removal, 2-D order extraction, and
wavelength calibration) was performed with a program developed and
extensively tested on exoplanet hosts by \citet{LarsB10}. For the fainter
stars, it was found preferable to divide the long exposures into three
pieces, and remove cosmic ray hits by median filtering.

Radial velocities were then derived from the reduced spectra by a
multi-order cross-correlation procedure. This operation is the most
difficult step in the analysis because {\it (i)} these stars are
extremely metal-poor and chemically peculiar; and {\it (ii)} the individual
spectra have low S/N ratios. Thus, selecting an optimum template spectrum
for each star is no trivial task. Noting that the primary objective of the
analysis is to measure small velocity {\it variations} rather than absolute
values, we have experimented with three types of template spectra: {\it (a)
} the highest S/N spectrum of each star; {\it (b)} the velocity-shifted and
co-added {\it mean} spectrum of each star, and {\it (c)} a synthetic
spectrum consisting of $\delta$ functions at the (solar) wavelengths of the
strongest visible lines. The choice of template for each star was then
guided by the consistency of the resulting velocities. Templates {\it (b)}
and {\it (c)} were generally found to give the most consistent results, the
latter also yielding velocities on a reliable absolute scale.

In summary, our final procedure yielded radial velocities with a standard
deviation of $\lesssim 100$ m s$^{-1}$ for the brighter and more metal-rich
stars, rising to $\sim 300-1000$ m s$^{-1}$ for the faintest stars with
the weakest spectral features.

\section{Binary Detection and Orbit Determination}

Fourteen of our stars exhibit no significant variation in radial velocity over
the period covered by our observations, including any earlier velocities
reported from HERES \citep{Barklem05}. Linear and parabolic fits of the run of
velocities vs. time were made to check for any long-term trends, but they were
generally negligible within the uncertainties; a few of the stars will be kept
under continued surveillance.

\begin{figure}
\plotone{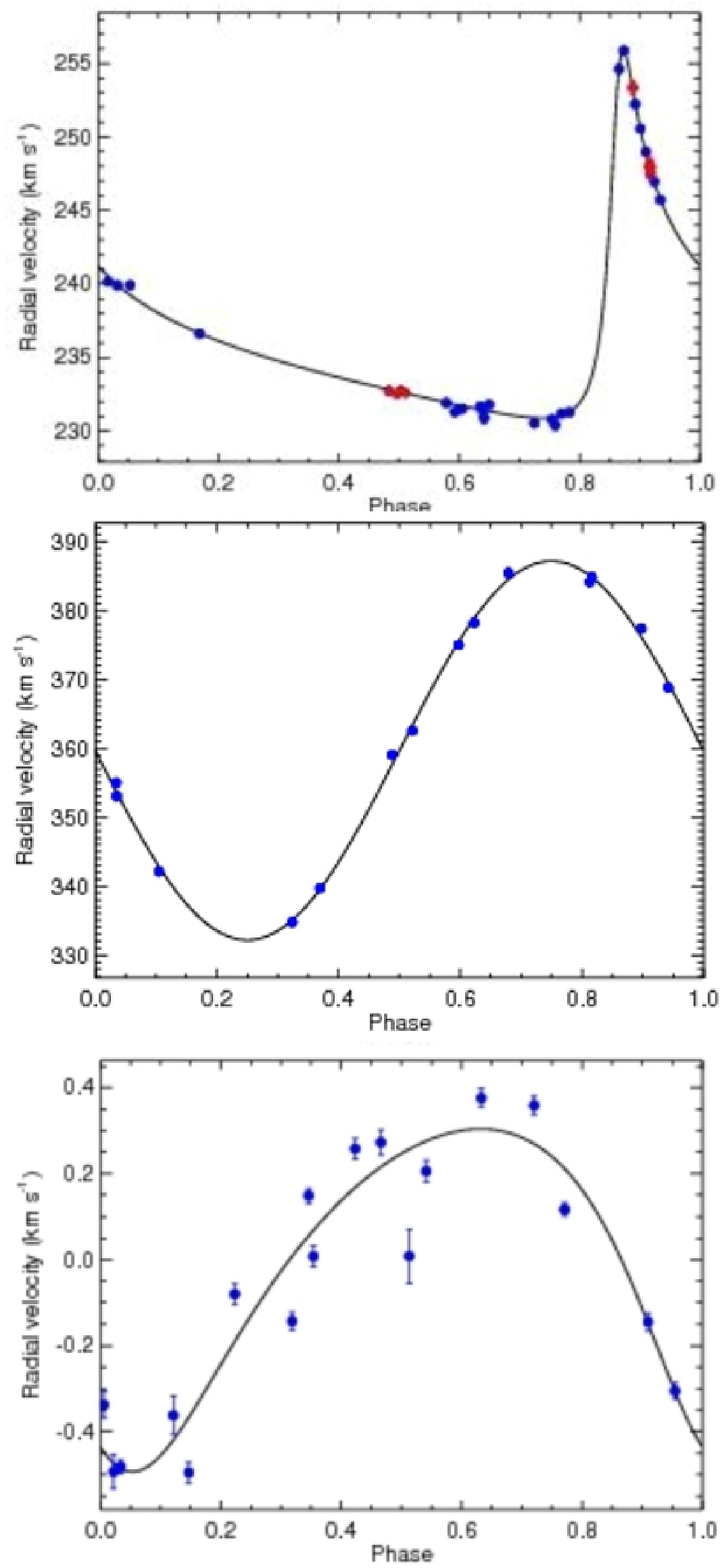} \caption{Observed and computed
spectroscopic orbits for HE~0442-1234 (top), HE~1044-2509 (middle)
and HE~1523-0901 (bottom). Orbital elements are listed in Table
\ref{tbl-orb}. {\it Blue dots:} FIES velocities; {\it red dots:}
Earlier data.\label{fig-orb}}
\end{figure}

The star HE~0442-1234 was shown by \citet{PB10} to be a probable
long-period spectroscopic binary. Our new data enabled us to complete the orbit
of this star, as well as for the newly-discovered binaries HE~1044-2509 and
HE~1523-0901. The orbital elements for these stars are listed in Table
\ref{tbl-orb}, and the observed and computed velocity curves are shown in Fig.
\ref{fig-orb}.

In general, the individual component masses cannot be derived
directly, but assuming a standard value of 0.8 $M_{\sun}$ for the
mass of a halo giant allows us to estimate a minimum mass for the
unseen companion. For HE~0442-1234, this is 0.67 $M_{\sun}$ if the
inclination $i \sim90\deg$, and it cannot be much larger for the
secondary star to remain invisible in the spectrum. In the other two
systems, a secondary star on the main sequence could be of similar
or lower mass. Assuming $M_2 = 0.6 M_{\sun}$ leads to the estimates
of $i$ given in Table \ref{tbl-orb}; note that the orbit of
HE~1523-0901 is seen nearly face-on. These, in turn, lead to
estimates of the size (volume equivalent radius) of the Roche lobes
of the unseen stars, which are not very sensitive to the adopted
geometry.

\label{sect_bin}

\begin{deluxetable}{lccc}
\tabletypesize{\scriptsize}
\tablecaption{Orbital Elements for the Detected Binary Stars\label{tbl-orb}}
\tablewidth{0pt}
\tablehead{
\colhead{Element} & \colhead{HE 0442-1234} & \colhead{HE 1523-0901} &
\colhead{HE 1044-2509}
}
\startdata
P (d)                            & 2513.38$\pm$4.46  & 302.78$\pm$0.78
                                &  36.57$\pm$0.20  \\
K (km s$^{-1}$)                 &   12.50$\pm$0.20  & 0.399$\pm$0.008
                                &   27.48$\pm$0.22  \\
$e$                             &  0.760$\pm$0.001  &  0.23$\pm$0.129
                                &    0.000$\pm$0.000  \\
$\gamma$ (km s$^{-1}$)           &  236.16$\pm$0.20  & 162.50$\pm$0.10
                                &  359.28$\pm$0.55      \\
$f(m)$ ($M_{\sun}$)             & 0.1396$\pm$0.0012 &
                                  1.59 $10^{-6}\pm$0.54 $10^{-6}$
                                & 0.0823$\pm$0.0027  \\
$a \sin i$ ($R_{\sun}$)         &    403.2$\pm$1.2  & 2.21$\pm$0.17
                                &  20.24$\pm$0.24    \\
$i$ (for $M_2 \sim 0.6 M_{\sun}$) &     88  & 1.5: & 65   \\
$R_{Roche}$ ($R_{\sun}$)        &   65\tablenotemark{a}  &  48  & 13    \\
$\sigma$ (1 obs., km s$^{-1}$)  &     0.28  & 0.11 &  0.98 \\
\enddata
\tablenotetext{a}{Size at periastron.}
\end{deluxetable}

\section{Discussion}
\label{sect_disc}

Our results enable us to address several issues of importance for
understanding the likely astrophysical site(s) of the $r$-process, and also
shed some light on the early Galactic production of carbon, as described
below.

\subsection{Binary Frequency}

With three binaries detected in a sample of 17 stars, the binary frequency of
the {\it r}-process-enhanced stars is $18$\%. The star HE~2327-5642, an
r-II star showing a possibly variable velocity \citep{Mash10}, is below the
southern limit of our sample, but is another candidate. This is fully consistent
with the $\sim$20\% binary frequency determined for normal cluster giants by
\citet{Merm96}. An additional 1-2 future discoveries in a larger sample might
boost the frequency to perhaps as much as 25\%, but there is clearly no evidence
that {\it all} {\it r}-process-element enhanced stars are binaries, as speculated by
\citet{QW01} and others.

Within the limitations imposed by the small sample, there also seems to be no
difference in the binary population of the r-I and r-II classes. Similarly, of
the two stars with measured U abundances, CS~31082-001 exhibits the so-called
``actinide boost" (an over-abundance of Th and U relative to third-peak {\it
r}-process elements such as Eu) and is a single star, while HE~1523-0901 is a
binary and shows no actinide boost. Remarkably, the C-enhanced prototypical r-II
star CS~22892-052 also seems {\it not} to be a binary, despite the earlier
suggestion by \citet{PreSne01}, indicating that its C content was not produced
in a former AGB companion. Thus, membership in a binary system appears to be
decoupled from details associated with these particular abundance variations.

\subsection{Orbital Properties}

The periods and eccentricities of our three confirmed binaries are entirely
consistent with those of the sample of chemically normal Population I giant
binaries by \citet{Merm96}: The longest-period orbit is highly eccentric,
and the shortest-period orbit is well below the limit of $\sim 100$d for
tidal circularization. Note that the old, metal-poor CH stars typically
have circular orbits for periods up to $\sim 1000$d. Moreover, the
secondary Roche lobes are too small to accommodate typical AGB stars of
$\sim$ 200 $R_{\sun}$, if the secondary stars passed through this phase of
their evolution. This is again consistent with the lack of {\it
s}-process-element enhancements in these stars.

\subsection{The Astrophysical Site of the {\it r}-Process}

Proving that the {\it r}-process-element production did not occur in
binaries provides no direct evidence of the nuclear physics process at the
production site above that afforded by the detailed abundance analyses of
these stars. Models attempting to explain the observed {\it r}-process
abundance patterns fall into two classes, core-collapse supernovae (SNe)
and merging binary neutron stars \citep[see, e.g.,][]{argast04,Gor11}, also
discussed comprehensively by \citet{sneden08}, and in the context of
CS~31082-001 by \citet{LP15}.

The key feature to be highlighted here is that the newly-formed {\it
r}-process elements were added in variable, but {\it internally consistent}
proportions, in the otherwise constant chemical composition of the next
generations of EMP and VMP stars. That the {\it r}-process-element
abundances varied so strongly in the clouds from which these stars formed
indicates that the {\it r}-process elements were not simply uniformly
dispersed, together with all lighter elements, in the supernova explosions
that are a common feature of all astrophysical models for the {\it
r}-process. Ejection in a jet or beam directly from the nascent neutron
star(s) seems the most natural scenario for achieving this. The chemical
composition of the progenitor, and the varying distance and direction of
the jet from the next cloud, would then explain, in a natural way, the
continuously varying proportions of {\it r}-process elements to the bulk
composition of the following generation of EMP stars.

\subsection{{\it r}-Process-Enhanced Stars as Chemical Tags of the Early
Galactic ISM}

In this scenario, most stars would receive a ``standard" dose of {\it
r}-process elements; stars with {\it r}-process-element abundances exceeding the
average by a factor of $\gtrsim2-3$ would be seen as {\it r}-process rich, while
those below the average \citep[exemplified by HD 122653,][]{westin00,
Honda06} would have been bypassed by the {\it r}-process ejection and
appear as {\it r}-process {\it poor}. This latter group may be as numerous
as the former, but without spectacularly strong spectral lines to call
attention to them. The recent discovery by \citet{aoki10} of a cool, EMP
main-sequence dwarf with highly {\it r}-process-enhanced elemental
abundance ratios, consistent with classification as an r-II star, would
obviate any model to explain {\it r}-process enhancements as only due to
some atmospheric anomaly confined to red giants.

The existence of {\it r}-process-element enhanced (or depleted) stars in a
narrow range of metallicity near [Fe/H] $\sim -3$ would imply that such
anisotropic SN explosions only appeared at a certain ``chemical time'', and
that the ISM was quickly fully mixed soon thereafter. The spectacular abundance
anomalies of these stars can thus be used as extreme examples of ``chemical
tags" of the sites and times of their formation. The r-I/r-II
classification is just a coarse tool to indicate the degree of enhancement.
However, as r-I stars are also found at higher metallicity than the r-II
stars, they may have formed from clouds that were further enriched by
material of ``normal'' halo composition.

The different processes responsible for the light and heavy {\it r}-process
elements \citep[see, e.g.,][]{LP8,Montes07}, as well as the existence of stars with
and without an actinide boost, remain to be explained by further modeling,
but the binary properties of EMP stars apparently played no role in this
context: Binaries formed with similar properties as in chemically normal
stars
\citep[e.g.,][]{GonHer08}.

Finally, it is remarkable that the prototypical r-II star CS~22892-052 is
single {\it and} significantly enhanced in carbon, which has been assumed to
originate in long-period binary stars together with the {\it s}-process
elements, which are {\it not} observed in CS~22892-052. This casts doubt
on the accepted explanation for the synthesis of C in the early Galaxy as
due primarily to pollution by former AGB binary companions, and suggests
the synthesis of C, N, and O in earlier, rapidly rotating massive stars as
one attractive alternative \citep[see, e.g., ][]{Meynet06,Meynet10}.

\section{Conclusions}
\label{sect_concl}

Eighty percent of our program r-I and r-II stars exhibit no detectable
radial-velocity variations, while three stars are binaries with well-determined
orbits (Table \ref{tbl-orb}), typical of systems with giant primaries, but no
AGB secondary stars. Thus, the binary population among these stars is normal,
and binary stars play no special role in producing the {\it r}-process elements
and injecting them into the early ISM. The case of CS~22892-052 suggests
that this may be true for the early synthesis of carbon as well.

We conclude that whatever progenitors produced the {\it r}-process elements
(and carbon) were {\it extrinsic} to the EMP and VMP stars we observe
today. These elements were likely ejected in a collimated manner, and make
these stars archetypical chemical indicators of their formation sites
in the early Galaxy.

\acknowledgments

This paper is based on observations made with the Nordic Optical Telescope,
operated on the island of La Palma jointly by Denmark, Finland, Iceland,
Norway, and Sweden, in the Spanish Observatorio del Roque de los Muchachos
of the Instituto de Astrofisica de Canarias. We thank Drs. Piercarlo
Bonifacio, Luca Sbordone, Lorenzo Monaco, and Jonay Gonzalez Hernandez for
alerting us to the binary nature of HE~0442-1234 and for allowing us to
include their velocities in our orbital solution, Dr. G. Torres for computing the orbit of HE1523-091, and Dr. Paul Barklem for providing the observing dates of the HERES spectra. We also thank several NOT
students for obtaining most of the NOT observations for us in service mode.
J.A. and B.N. acknowledge support from the Danish Natural Science Research
Council, and L.A.B from the Carlsberg Foundation.
T.C.B. acknowledges partial funding of this work from grants PHY
02-16783 and PHY 08-22648: Physics Frontier Center/Joint Institute for
Nuclear Astrophysics (JINA), awarded by the U.S. National Science
Foundation.



\end{document}